\documentclass{WileyMSP-template}
\usepackage{amsmath}
\usepackage{amssymb}
\usepackage[numbers,sort,compress]{natbib}
\usepackage{color}
\begin{document}
\pagestyle{fancy}
\rhead{\includegraphics[width=2.5cm]{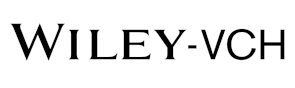}}
\title{Stable high-dimensional weak-light soliton molecules and their active control}
\maketitle\\
\author{Lu Qin}
\author{Chao Hang}
\author{Boris A. Malomed}
\author{Guoxiang Huang*}
\begin{affiliations}\\
Dr. L. Qin, Prof. C. Hang, Prof. G. Huang\\
State Key Laboratory of Precision Spectroscopy,
East China Normal University, Shanghai 200062, China\\
Prof. C. Hang, Prof. G. Huang\\
NYU-ECNU Institute of Physics, New York University at Shanghai, Shanghai 200062, China\\
Collaborative Innovation Center of Extreme Optics, Shanxi University, Taiyuan, Shanxi 030006, China\\
Email:gxhuang@phy.ecnu.edu.cn\\
Dr. L. Qin,\\
Department of Physics, Henan Normal University, Xinxiang 453007, China\\
Prof. B. A. Malomed\\
Department of Physical Electronics, School of Electrical Engineering, Faculty of Engineering, Tel Aviv University, Tel Aviv, Israel\\
Instituto de Alta Investigaci\'{o}n, Universidad de Tarapac\'{a}, Casilla 7D, Arica, Chile
\end{affiliations}

\begin{abstract}
Bound states of solitons, alias \textit{soliton molecules} (SMs), are well known in one-dimensional (1D) systems, while making stable bound states of multidimensional solitons is a challenging problem because of the underlying instabilities. Here we propose a scheme for the creation of \textit{stable} (2+1)D and (3+1)D optical SMs in a gas of cold Rydberg atoms, in which electromagnetically induced transparency (EIT) is induced by a control laser field. We show that, through the interplay of the EIT and the strong long-range interaction between the Rydberg atoms, the system gives rise to giant nonlocal Kerr nonlinearity, which in turn supports stable (2+1)D spatial optical SMs, as well as ring-shaped soliton necklaces, including rotating ones. They feature a large size, low generation power, and can be efficiently manipulated by tuning the nonlocality degree of the Kerr nonlinearity. Stable (3+1)D spatiotemporal optical SMs, composed of fundamental or vortex solitons, with low power and ultraslow propagation velocity, can also be generated in the system. These SMs can be stored and retrieved through the switching off and on of the control laser field. The findings reported here suggest applications to data processing and transmission in optical systems.
\end{abstract}

\section{Introduction}
Solitons are self-trapped wave packets maintained by the interplay between dispersion (and/or diffraction) and nonlinearity of host media \cite{Zakharov,Ablowitz}. They are ubiquitous in nature, having been discovered
in many areas, including hydrodynamics and plasmas \cite{Rubenchik}, optics
\cite{Trillo2001,Hasegawa2002,Kivshar2002,Akhmediev2005,Malomed2006,Lederer,Wadati,Maimistov,Turitsyn,
Grelu2012,Assanto2013}, Bose-Einstein condensates \cite{Konotop,Abdullaev,DFrantz,Salasnich}, superconductivity \cite{Ustinov-book}, solid-state physics, magnetic media, etc.~\cite{Dauxois-Peyrard,Kevrekidis-book,Barcelona}. While in (nearly) integrable systems solitons interact elastically~\cite{Zakharov,Ablowitz}, collisions
between them in nonintegrable settings exhibit a variety of outcomes, including, in particular, fusion, fission, and annihilation \cite{Tappert1971,RMP,Tikhonenko1996,Krolikowski1997,Krolikowski1998,Herrmann,Lu2004,
Pfeiffer2006,Randy,Descalzi2009,Maiden2014,Kottig2017,Descalzi2020}. In this context, bound states of solitons
\cite{Malomed1991,Malomed1993,Nepom,random,Afanasjev1997,Tang-1,Tang-2,Komarov2009,Wang2017,Skryabin,Gelash2019},
often referred as soliton molecules (SMs), are objects of great interest, as they demonstrate unique properties and offer various potential applications for working with mode-locked fiber lasers, matter waves, optical microresonators, polariton superfluids, and other physical realizations~\cite{Akhmediev1997,Crasovan2003,Vekslerchik2003,Stratmann2005,
Hause2008,Leblond,Zavyalov2009,Khawaja2010,Yin2011,Khawaja2011,Santos2012,
Rohrmann2013,Boudje2013,Hause2013,Shirley2014,Alamoudi2014,Turmanov2015,Baizakov2015,
Leblond-2,Wang2016,Herink2017,Krupa2017,YiyangLuo2017,XuemingLiu2018,Qin2018,
JunsongPeng2018,GangXu2019,Melchert2019,Wang2019,ZhouY2020,Kurtz2020,Weng2020,
Maitre2020}. Especially promising applications of SMs in the area of photonics include the design of new laser schemes, switchers, and data carriers~\cite{Yin2011,Rohrmann2013,Alamoudi2014,Kurtz2020}.

Previous studies of SMs were chiefly limited to one-dimensional (1D) systems with local nonlinearity, such as temporal SMs in fiber lasers.
In such systems, the interaction between solitons in the SM is determined by the overlap of ``tails" of the wave functions of adjacent solitons, which rapidly decays as the separation between the solitons increases. As a result, the size of SMs in locally nonlinear media usually does not exceed two or three widths of the single soliton, and they may be readily subject to instability against long-wavelength transverse perturbations when the 1D setting is embedded in the 3D space. In fact, unlike 1D solitons, which are normally stable states, stability of 2D and 3D solitons is a problem as the usual cubic self-focusing gives rise to the critical and supercritical collapse in the 2D and 3D space, respectively \cite{Rubenchik,Berge',Sulem,spatiotemp,Fibich}, which makes multidimensional solitons unstable in these simple models. Still more unstable, against spontaneous splitting, are 2D and 3D solitons with
embedded vorticity \cite{Michinel,Pego,vortices}. Therefore, identification of physically relevant mechanism for stabilizing fundamental and vortex solitons is a problem of fundamental significance. In recent years, it was addressed in various models, leading to predictions of stable multidimensional solitons in relatively sophisticated settings \cite{spatiotemp,vortices,Special-topics,Astra,Kengne}. Experimentally, stable 2D spatial solitons
were created in an optical medium with cubic-quintic nonlinearity \cite{Cid}, and in various forms of quasi-soliton ``quantum droplets" in BECs \cite{Leticia,Leticia2,Inguscio,hetero,Pfau,droplets}. As for 2D solitons with embedded vorticity, so far they were experimentally demonstrated only in a transient form, being temporarily stabilized in an optical material with the saturable self-focusing and three-photon absorption \cite{Cid2}.

Another promising possibility is the use of optical media with nonlocal interactions, which make it possible to support stable multidimensional solitons \cite{Snyder1997,Krolikowski2000,Conti2004,Xu2005,Alberucci2006,Buccoliero2007,Skupin2007,Burgess2009,Conti2010,Maucher2011,Sevincli2011,
Lahav2017,Horikis2017,Li2018,Wilson2018,Bai2019} and long-range interactions between them \cite{Rotschild2006,Vardi}. The objective of the present work
is to elaborate a scenario for the creation of stable 2D and 3D optical SMs in a cold Rydberg atomic gas, under the condition of the electromagnetically induced transparency (EIT)~\cite{Fleischhauer2005}. The large electric-dipole moments of Rydberg atoms give rise to strong long-range interactions between them~\cite{Gallagher2008,Saffman2010,Adams2020}. The interplay of Rydberg-Rydberg interactions with EIT gives rise to a giant nonlocal Kerr
nonlinearity~\cite{Firstenberg2016,Murray2016,Bai2016,Adams2017}.
In this work, we find that the system supports stable 2D spatial optical SMs that can be built of fundamental or vortex solitons, with the size more than six times the soliton's width
($\sim 100\,\mu \mathrm{m}$). Furthermore, the power required to generate SMs is found to be at
$\sim \mu$W, which is, at least, three orders of magnitude smaller than SM-generation power in fiber-laser systems and solid-state media. Such as lead glass, where it may be up to several watts~\cite{XuemingLiu2018,Rotschild2006,Wang2019,Herink2017,Krupa2017,YiyangLuo2017}; moreover, properties of the Rydberg-EIT medium make it possible to efficiently control the size of SMs by tuning the degree of the nonlocality of the Kerr nonlinearity.
In addition to two-soliton molecules, stable ring-shaped bound states built of several solitons (``necklaces"), fundamental or vortex ones, carrying overall
phase circulation, are constructed too, including rotating necklaces.

We also find that stable bound states of fundamental and vortex
3D spatial-temporal solitons, with ultralow propagation
velocity ($\sim 10^{-5}\,c$, where $c$ is the light speed in vacuum) and
ultralow generation power, can be created by means of an appropriate
combination of nonlocal and local Kerr nonlinearities in such a Rydberg gas.
 An essential asset of the system under the consideration is that it is highly controllable, admitting one to
store and retrieve the predicted 3D spatiotemporal SMs with high fidelity by
switching the control field off and on. The results reported here
suggest the realization of novel SMs at weak-power levels, and
implementation of their effective manipulation, with potential applications to
optical data processing and transmission.

The following presentation is arranged as follows. In Sec.~II, we describe
the physical model and derive an envelope equation governing the propagation
of the probe field. In Sec.~III, we address the interaction between spatial
solitons, formation of SMs and necklace-shaped bound states, and
manipulation with them by adjusting the nonlocal Kerr nonlinearity. The
possibilities to create and manipulate stable spatiotemporal SMs are reported in
Sec. IV. The work is summarized in Sec.~V.

\section{The model}

\subsection{The physical setup}

\begin{figure}[tbp]
\centering
\includegraphics[width=0.55\textwidth]{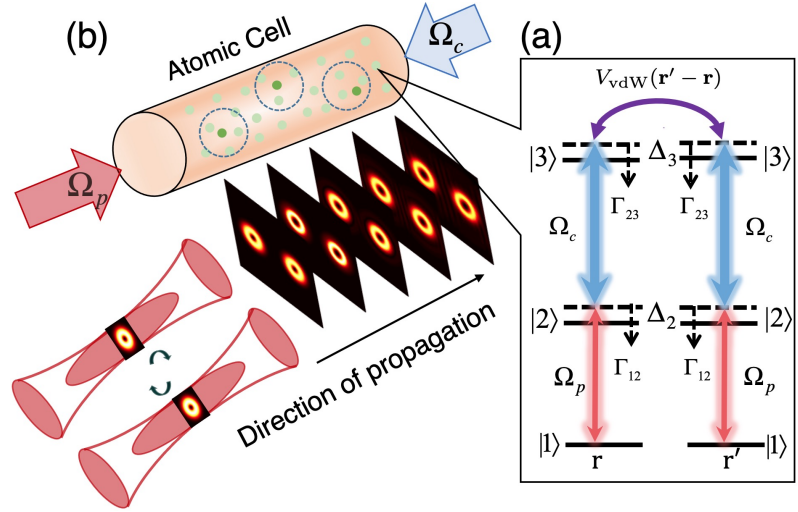}
\caption{{\protect\footnotesize Schematics of the model. (a)~Energy-level diagram and excitation scheme of
two ladder-type three-level atoms. $|1\rangle$, $|2\rangle$, and $|3\rangle$ are respectively the ground, intermediate, and Rydberg states;
$\Gamma _{12}$ and $\Gamma _{23}$ are respectively decay rates of $|2\rangle $ and $|3\rangle $; $\Omega _{p}$ and $\Omega_{c}$ are respectively half Rabi frequencies of the probe and control laser fields; $\Delta_2$ and $\Delta_3$ are respectively one- and two-photon detunings.
The interaction between the two Rydberg atoms is
described by the van der Waals potential $V_{\mathrm{vdW}}(\mathbf{r^{\prime
}}-\mathbf{r})=-\hbar C_{6}/|\mathbf{r^{\prime }}-\mathbf{r}|^{6}$.
(b)~Top part: possible experimental geometry. Bottom part: the
contactless interaction between two optical vortices, which form a stable vortex molecule.}}
\label{fig1}
\end{figure}

We start by considering a laser-cooled, dilute three-level atomic gas, interacting with a weak probe laser field $\mathbf{E}_{p}$ with center frequency $\omega _{p}$ (wavenumber $k_{p}=\omega_{p}/c $), driving the transition $|1\rangle \leftrightarrow |2\rangle $, and a strong, continuous-wave control laser field $\mathbf{E}_{c}$ with frequency $\omega _{c}$ (wavenumber $k_{c}=\omega_{c}/c$), driving the transition $|2\rangle\leftrightarrow |3\rangle $; see Fig.~\ref{fig1}(a).
Here $|1\rangle $, $|2\rangle $, and $|3\rangle $ denote, respectively, the ground, intermediate, and high-lying Rydberg states;
$\Gamma _{12}$ and $\Gamma _{23}$ are spontaneous-emission decay rates from $|2\rangle $ to $|1\rangle $ and from $|3\rangle $ to $|2\rangle $, respectively. The interaction between the two Rydberg atoms respectively at positions $\mathbf{r}$ and $\mathbf{r}^{\prime }$ is described by the van der Waals (vdW) potential
\begin{equation}
V_{\mathrm{vdW}}=\hbar V(\mathbf{r}^{\prime }-\mathbf{r})\equiv -\frac{\hbar
C_{6}}{|\mathbf{r^{\prime }}-\mathbf{r}|^{6}},  \label{vdW}
\end{equation}
with $C_6$ the dispersion coefficient~\cite{Gallagher2008,Saffman2010}.
The initial atomic population is prepared in the ground state $|1\rangle $. The total electric-field vector in the system is given by $\mathbf{E}=\mathbf{E}_{c}+%
\mathbf{E}_{p}\equiv \sum_{l=c,p}{\mathbf{e}_{l}\mathcal{E}}_{l}\exp [i(%
\mathbf{k}_{l}\cdot \mathbf{r}-\omega _{l}t)]+\mathrm{c.c}.$, where c.c.
stands for the complex conjugate, while $\mathbf{e}_{c}$ and $\mathbf{e}_{p}$
($\mathcal{E}_{c}$ and $\mathcal{E}_{p}$) are, respectively, polarization
unit vectors (envelopes) of the control and probe fields. To suppress the
Doppler effect, the probe and control fields are assumed to
counter-propagate along the $z$ direction, i.e., $\mathbf{k}_{p}=k_{p}%
\mathbf{e}_{z}$ and $\mathbf{k}_{c}=-k_{c}\mathbf{e}_{z}$, with
$\mathbf{e}_{z}$ being the unit vector of the $z$ direction.

The Hamiltonian of the system is given by $\hat{H}=\mathcal{N}_a\int d^3{r}\hat{\mathcal{H}}_0({\bf r},t)+(\mathcal{N}_a/2)\int d^3{r}\hat{\mathcal{H}}_1({\bf r},t)$. Here $d^3r=dxdydz$,
$\mathcal{N}_a$ is atomic density, $\hat{\mathcal{H}_0}({\bf r},t)$ is the Hamiltonian density describing the atoms and the coupling between the atoms and light fields, $\hat{\mathcal{H}_1}({\bf r},t)$ is the Hamiltonian density describing the Rydberg-Rydberg interaction. Under the electric-dipole and rotating-wave approximations, $\hat{\mathcal{H}_0}$ and $\hat{\mathcal{H}_1}$ have the forms
\begin{eqnarray}
&& \hat{{\mathcal H}}_0= -\sum_{\alpha =2}^{3}{\hbar \Delta _{\alpha }\hat{S}%
_{\alpha \alpha }\left( \mathbf{r},t\right) }-\hbar \left[ \Omega _{p}\hat{S}%
_{12}+\Omega _{c}\hat{S}_{23}+\mathrm{h.c.}\right], \nonumber \\
&& \hat{\mathcal{H}}_1=\mathcal{N}_{a} \int{d^3 {r}^{\prime}\hat{S}_{33}({\bf r}',t) \hbar V ({\bf r}^{\prime}-{\bf r}) \hat{S}_{33} ({\bf r},t )}. \label{Hamiltonian1}
\end{eqnarray}
Here $\Delta_{2}=\omega_{p}-(E_{2}-E_{1})/\hbar$ and $\Delta_{3}=\omega
_{p}+\omega_{c}-(E_{3}-E_{1})/\hbar $ are, respectively, the one- and
two-photon detunings, with $E_{\alpha}$ being the eigenvalue of the energy
of the state $|\alpha \rangle $; $\hat{S}_{\alpha \beta }\equiv \hat{\sigma}
_{\beta \alpha}\exp\{i[(\mathbf{k}_{\beta }-\mathbf{k}_{\alpha })\cdot
\mathbf{r}-(\omega_{\beta }-\omega_{\alpha }+\Delta_{\beta }-\Delta
_{\alpha})t]\}$ is the atomic transition operator~\cite{note100};
$\Omega _{p}={(\mathbf{e}_{p}\cdot \mathbf{p}_{12})\mathcal{E}_{p}}/(2\hbar )$
and $\Omega_{c}={(\mathbf{e}_{c}\cdot \mathbf{p}_{23})\mathcal{E}_{c}}%
/(2\hbar )$ are, respectively, half Rabi frequencies of the probe and
control fields, with $\mathbf{p}_{\alpha \beta }$ the electric-dipole matrix
elements associated with the transition $|\beta\rangle \leftrightarrow
|\alpha \rangle $.  The potential $V$ in the expression of $\hat{\mathcal{H}}_1$ is taken as per Eq. (\ref{vdW})~\cite{Gallagher2008,Saffman2010,Firstenberg2016,Murray2016}.

The atomic dynamics is governed by the Heisenberg equation of motion for the operators
$\hat{S}_{\alpha\beta}({\bf r},t)$, i.e. $i\hbar\frac{\partial}{\partial t}\hat{S}_{\alpha\beta}({\bf r},t)=[\hat{H}, \hat{S}_{\alpha\beta}({\bf r},t)]$. Taking expectation values on the both sides of this equation, we obtain the optical Bloch equation involving one- and two-body reduced density matrices, with the form
\begin{eqnarray}\label{Bloch0}
\frac{\partial\hat{\rho}}{\partial t}=-\frac{i}{\hbar} \left[{\hat{ H}_{0}},\hat{\rho}\right]-\Gamma\left[\hat{\rho}\right]+\hat{R}\,[\hat{\rho}_{\rm 2body}],
\end{eqnarray}
where $\hat{\rho} ({\bf r},t)$ is reduced one-body  density matrix (DM) with matrix elements  $\rho_{\alpha\beta}({\bf r},t)\equiv\langle \hat{S}_{\alpha\beta} ({\bf r},t)\rangle$, $\Gamma$ is a $3\times 3$
relaxation matrix describing the spontaneous emission and dephasing. Due to the existence of the Rydberg-Rydberg interaction, two-body reduced DM, i.e. $\hat{\rho}_{\rm 2body}(\bf r^{\prime},{\bf r},t)$  [with DM elements $\rho_{\alpha\beta,\mu \nu} ({\bf r}',{\bf r},t)$], is involved in Eq.~(\ref{Bloch0}), represented by the last term $\hat{R}\,[\hat{\rho}_{\rm 2body}]$. The explicit form of Eq.~(\ref{Bloch0}) is given in Sec. 1 of Supporting Information.

From Eq.~(\ref{Bloch0}), we see that to get the solution of one-body DM elements $\rho _{\alpha \beta }$, equations for two-body elements $\rho _{\alpha \beta ,\mu \nu }$ are needed, which, in turn, involve
three-body DM element $\rho _{\alpha \beta ,\mu \nu ,\gamma \delta }$, and so on. As a result, one obtains a hierarchy of infinite equations for $N$-body DM elements ($N=1,\,2,\,3,\cdots $) that must be solved
simultaneously. To solve such a chain of equations, a suitable treatment beyond mean-field approximation must be adopted. A powerful one is the reduced density-matrix expansion, by which
the hierarchy of the infinite equations is truncated consistently and the
problem is reduced to solving a closed system of equations for the one- and
two-body DM elements, as elaborated recently~\cite{Bai2019,Bai2016,Mu2021}.

The propagation of the probe field is governed by the Maxwell equation,
which, under the slowly-varying-envelope approximation, is reduced to~\cite{Bai2016}
\begin{equation}
i\left( \frac{\partial }{\partial z}+\frac{1}{c}\frac{\partial }{\partial t%
}\right) \Omega _{p}+\frac{c}{2\omega _{p}}\nabla _{\perp }^{2}\Omega
_{p}+\kappa _{12}\,\rho _{21}=0.  \label{Max}
\end{equation}
Here $\nabla_{\perp }^{2}=\partial_{x}^{2}+\partial_{y}^{2}$ and $\kappa
_{12}=\mathcal{N}_{a}\omega_{p}|\mathbf{p}_{12}|^{2}/(2\epsilon_{0}c
\hbar)$.

\subsection{The nonlinear envelope equation}

Since the probe field is much weaker than the control field, the depletion of
the atomic population in the ground state is small and a standard
perturbation method can be applied to solve the system of Maxwell-Bloch (MB)
equations~(\ref{Bloch0}) and (\ref{Max}). To include the many-body
correlations produced by the strong Rydberg-Rydberg interactions in a reasonable way,
a beyond mean-field approximation~\cite{Bai2019,Bai2016,Mu2021} mentioned above
must be used. Then, in the leading-order
approximation, we obtain the solution for the probe field $\Omega
_{p}=F(x,y,z,t)e^{iK(\omega )z-i\omega t}$, where $F$ denotes a slowly
varying envelope function and $K(\omega )$ stands for the linear dispersion
relation, $K(\omega )=\omega /c+\kappa _{12}(\omega
+d_{31})/[|\Omega _{c}|^{2}-(\omega +d_{21})(\omega +d_{31})]$.

At the third-order approximation (details are given in Sec. 2 of the Supporting Information),
we obtain the nonlinear equation for the probe-field envelope
\begin{align}
& i\frac{\partial }{\partial z}\Omega _{p}+\frac{c}{2\omega _{p}}\nabla
_{\bot }^{2}\Omega _{p}-\frac{K_{2}}{2}\frac{\partial ^{2}}{\partial T^{2}}%
\Omega _{p}+W{\left\vert \Omega _{p}\right\vert ^{2}}\Omega _{p}+{\Omega _{p}}\iint dxdyG(x^{\prime }-x,y^{\prime }-y){\left\vert \Omega
_{p}(x^{\prime },y^{\prime },z,T)\right\vert }^{2}=0,
\label{NLS}
\end{align}%
where $T=t-z/V_{g}$, with $V_{g}=(\partial K/\partial \omega )^{-1}$ being
the group velocity of the envelope, and $K_{2}=\partial ^{2}K/\partial
\omega ^{2}$ defines the group-velocity dispersion. The last two terms in
Eq.~(\ref{NLS}) contributed, respectively, from the local and nonlocal optical Kerr nonlinearities in the system. Explicit expressions of the coefficients $W$ and $G$ (nonlocal nonlinear response function) are given in Sec. 2 of the Supporting Information. The local optical Kerr nonlinearity is contributed to by the short-range interaction between photons and atoms, proportional to the atomic density $\mathcal{N}_{a}$, whereas the nonlocal optical Kerr nonlinearity is contributed by the long-range Rydberg-Rydberg interaction, which scaled quadratically with the atomic density (i.e., proportional to $\mathcal{N}_{a}^2$).

Equation (\ref{NLS}) can be further written into the non-dimensional form
\begin{align}
& i\frac{\partial u}{\partial \zeta }+\left( \frac{\partial ^{2}}{\partial
\xi ^{2}}+\frac{\partial ^{2}}{\partial \eta ^{2}}\right) u+d\frac{\partial
^{2}u}{\partial \tau ^{2}}+w|u|^{2}u +u\iint d\xi ^{\prime }d\eta ^{\prime }g(\xi ^{\prime }-\xi ,\eta ^{\prime
}-\eta )\left\vert u\left(\xi^{\prime},\eta^{\prime},\zeta,\tau\right) \right\vert
^{2}=0,  \label{NLS1}
\end{align}%
where $u=\Omega _{p}/U_{0}$, $\zeta =z/(2L_{\mathrm{diff}})$, $(\xi ,\eta
)=(x,y)/R_{0}$, $\tau =T/\tau _{0}$, $d=-\mathrm{sgn}(K_{2})L_{\mathrm{diff}%
}/L_{\mathrm{disp}}$, $w=2L_{\mathrm{diff}}|U_{0}|^{2}W$, and $g=2L_{\mathrm{%
diff}}R_{0}^{2}|U_{0}|^{2}G(\xi ^{\prime }-\xi ,\eta ^{\prime }-\eta )$, with the diffraction and dispersion lengths respectively given by $L_{\mathrm{diff}}=\omega_{p}R_{0}^{2}/c$ and $L_{\mathrm{disp}}=\tau_{0}^{2}/|K_{2}|$. Here $U_{0}$,  $R_{0}$, and $\tau _{0}$ are typical half Rabi frequency, beam radius in the transverse plane ($x$,$y$), and temporal duration of the probe-field envelope, respectively.

To address a typical example, we consider laser-cooled strontium $^{88}%
\mathrm{Sr}$ atoms, with atomic levels $|1\rangle =\left\vert
5s^{2}~^{1}S_{0}\right\rangle$, $|2\rangle =\left\vert
5s5p~^{1}P_{1}\right\rangle$, and $|3\rangle =\left\vert
5sns~^{1}S_{0}\right\rangle$. For the principal quantum number $n=60$, the
dispersion parameter $C_{6}\approx 2\pi \times 81.6\,\mathrm{GHz}\cdot
\mu \mathrm{m}^{6}$ (which implies the Rydberg-Rydberg interaction is attractive). The spontaneous emission decay rates are $\Gamma_{12}\approx 2\pi \times 32$\,MHz and $\Gamma_{23}\approx 2\pi\times 16.7$\,kHz, while the detunings are taken to be $\Delta _{2}=-2\pi\times 240$\,MHz and $\Delta_{3}=-2\pi\times 0.16$\,MHz. The density of the atomic gas is $\mathcal{N}_{a}=9\times 10^{10}~\mathrm{cm}^{-3}$, and the half Rabi frequency of the control field is
$\Omega_{c}=2\pi \times 5$\,MHz. Since $\Delta _{2}\gg \Gamma _{12},\,\Delta _{3}$, which makes the system works in a dispersive nonlinearity regime,
the imaginary parts of coefficients in Eq.~(\ref{NLS}) are much smaller than
the corresponding real parts, and hence Eq.~(\ref{NLS1}) can be approximately
considered as a real-coefficient one.

The nonlocal nonlinear response function $g(\xi^{\prime }-\xi,
\eta^{\prime }-\eta )$ has a very complicated expression. For the convenience of
the subsequent variational calculation for the interaction force between
solitons, we approximate it by a Gaussian function, i.e.,
\begin{equation}
g\approx g_{F}
\equiv \frac{g_{0}}{(0.64\,\sigma\sqrt{\pi })^{2}}\,e^{-\frac{(\xi-\xi')^2
+(\eta-\eta')^{2}}{(0.64\,\sigma)^{2}} }, \label{Gauss}
\end{equation}
(details of relations between $g$ and $g_{F}$ are given in Sec. 3 of the Supporting Information), where $g_{0}=\iint d\xi d\eta g(\xi ^{\prime }-\xi
,\eta-\eta)$ is a constant and $\sigma $ characterizes the
nonlocality degree of the nonlinearity, defined as
\begin{equation}
\sigma =R_{b}/R_{0}.  \label{sigma}
\end{equation}
Here $R_{b}=(|C_{6}/\delta _{\mathrm{EIT}}|)^{1/6}$ denotes the Rydberg
blockade radius, where $\delta _{\mathrm{EIT}}\approx |\Omega
_{c}|^{2}/|\Delta _{2}|$ is the linewidth of the EIT transmission spectrum
for $|\Delta _{2}|\gg \Gamma _{12}$~\cite{Firstenberg2016,Murray2016}. With
the values of parameters adopted above, we get
$R_{b}\approx 9.6\,\mu\mathrm{m}$.
If $R_{b}\ll R_{0}$ (i.e., $\sigma \rightarrow 0$, the local limit), the
nonlocal response reduces to the delta function, i.e., $g_{0}\delta (\xi
-\xi ^{\prime },\eta -\eta ^{\prime })$. In this limit, the nonlocal Kerr
nonlinearity is reduced to the usual local term, $g_{0}|u|^{2}u$. If $%
R_{b}\gg R_{0}$ ($\sigma \rightarrow \infty $, the strongly nonlocal limit),
the nonlocal response, given by Eq.~(\ref{Gauss}), reduces to a linear term $%
g_{0}Pu$, where $P=\iint d\xi d\eta \,|u|^{2}$ is the power of the probe
field (this limit corresponds to the so-called \textquotedblleft
accessible-soliton''  model, which is actually a limit
one~\cite{Snyder1997,Conti2004}).

The susceptibility of the probe field is defined by $\chi _{p}=\mathcal{N}%
_{a}(\mathbf{e}_{p}\cdot \mathbf{p}_{12})^{2}\rho _{21}/(\varepsilon
_{0}\hbar \Omega _{p})$, which can be further expressed as $\chi _{p}\approx
\chi ^{(1)}+\chi _{\mathrm{loc}}^{(3)}|\mathcal{E}_{p}|^{2}+\chi _{\mathrm{%
nloc}}^{(3)}|\mathcal{E}_{p}|^{2}$. In this expansion, $\chi ^{(1)}$
represents the linear susceptibility; $\chi _{\mathrm{loc}}^{(3)}$ and
$\chi _{\mathrm{nloc}}^{(3)}$ are respectively the local and nonlocal
third-order nonlinear susceptibilities, associated to the coefficients
of Eq.~(\ref{NLS}) as
\begin{align}
& \chi _{\mathrm{loc}}^{(3)}=\frac{2(\mathbf{e}_{p}\cdot \mathbf{p}_{12})^{2}%
}{k_{p}\hbar ^{2}}W, \quad \chi _{\mathrm{nloc}}^{(3)}=\frac{2(\mathbf{e}_{p}\cdot \mathbf{p}%
_{12})^{2}}{k_{p}\hbar ^{2}}\iint dx^{\prime }dy^{\prime }G(x^{\prime
}-x,y^{\prime }-y).
\end{align}%
With the system's parameters introduced above, we obtain $\chi _{\mathrm{loc}%
}^{(3)}\sim 10^{-11}~\mathrm{m^{2}V^{-2}}$ and $\chi _{\mathrm{nloc}%
}^{(3)}\sim 10^{-8}~\mathrm{m^{2}V^{-2}}$, i.e., the nonlocal optical Kerr
nonlinearity is three orders of magnitude stronger than the local one due to the Rydberg-Rydberg interaction.

\section{Nonlocal (2+1)D spatial solitons and vortex molecules}

\subsection{The interaction between two nonlocal (2+1)D spatial solitons}

We now address the interaction force between two nonlocal
(2+1)D~\cite{note101} spatial solitons. By setting $d=0$ (which implies that time
duration $\tau _{0}$ of the probe field is large enough to make the
dispersion of the system negligible, valid for $L_{\mathrm{disp}
}\gg L_{\mathrm{diff}}$), Eq.~(\ref{NLS1}) reduces to the form
\begin{eqnarray}
&& i\frac{\partial u}{\partial \zeta}+\left(\frac{\partial^{2}}{\partial
\xi^{2}}+\frac{\partial^{2}}{\partial \eta ^{2}}\right) u+w|u|^{2}u
+u\iint d\xi^{\prime }d\eta^{\prime }g(\xi^{\prime }-\xi ,\eta ^{\prime
}-\eta )\left\vert u\left(\xi^{\prime},\eta^{\prime},\zeta\right)
\right\vert^{2}=0.
\end{eqnarray}
The Lagrangian corresponding to this equation is $L=\iint_{-\infty
}^{\infty }d\xi d\eta \,\mathcal{L}$, with the density
$\mathcal{L}=\frac{i}{2}(uu_{\zeta}^{\ast}-u^{\ast }u_{\zeta})+|u_{\xi
}|^{2}+|u_{\eta}|^{2}-\frac{w}{2}|u|^{4}
-\frac{1}{2}|u|^{2}\iint g(\xi -\xi ^{\prime},\eta -\eta ^{\prime
})\left\vert u\left(\xi ^{\prime},\eta ^{\prime},\zeta\right) \right\vert
^{2}d\xi d\eta$.

The bound state of two solitons (i.e., the two-soliton \textquotedblleft
molecule'') is sought by means of the ansatz
\begin{equation}
u(\xi ,\eta )=u_{+}(\xi ,\eta )-u_{-}(\xi ,\eta ),\label{+-}
\end{equation}%
with each soliton approximated by a Gaussian, i.e.,
\begin{equation}
u_{\pm }=A\,e^{-[\left(\xi \pm \mathcal{D}/{2}\right)^{2}+\eta^{2}]/(2a^{2})
+ib(\xi^{2}+\eta ^{2})+i\phi}.  \label{ansatz0}
\end{equation}
It includes two identical Gaussian beams, located at different positions
$(-\mathcal{D}/2,0)$ and $(\mathcal{D}/2,0)$, with a $\pi $ phase difference.
Variational $z$-dependent parameters are $A$ (amplitude), $\mathcal{D}$ (spatial separation), $b$ (chirp), and $\phi$ (phase), while the total power $P=2\pi a^{2}A^{2}[1-e^{-\mathcal{D}^{2}/(4a^{2})}]$ is a conserved quantity.
Due to the $\pi $ phase difference, there is a repulsive interaction between the two solitons. Such a repulsive interaction is expected to balance the attractive interaction induced by the self-focusing nonlocal optical Kerr nonlinearity (induced by the attractive Rydberg-Rydberg interactions), and thus help to form a stable two-soliton molecule~\cite{potential}.

Following the standard procedure of the variational approximation~\cite{Bai2019,potential,Raghavan2000,VA}, one can derive evolution equations for the variational parameters $A$, $\mathcal{D}$, $b$, and $\phi$, and hence the equation of motion for the spatial separation $\mathcal{D}$ between center-of-mass positions of the interacting solitons
\begin{equation}
\frac{d\mathcal{D}}{d\zeta }=\frac{16b\left[1-e^{-\mathcal{D}^{2}/(4a^{2})}\right]\left[(a^{2}+%
\mathcal{D}^2/4)e^{\mathcal{D}^{2}/(4a^{2})}-a^{2}\right]}{\mathcal{D}[(e^{\mathcal{D}%
^{2}/(4a^{2})}-1)-\mathcal{D}^{2}/(4a^{2})]}. \label{VA1}
\end{equation}
This equation can be cast in the form of the equation of motion for the
Newtonian particle $M_{s}(d^{2}\mathcal{D}/d\zeta ^{2})=-\partial U/\partial \mathcal{D}$,
where $M_{s}$ is the effective mass of each soliton and $U=U(\mathcal{D})$
denotes the effective potential which accounts for the interaction between
the two solitons.

\begin{figure}[tbp]
\centering
\includegraphics[width=0.65\columnwidth]{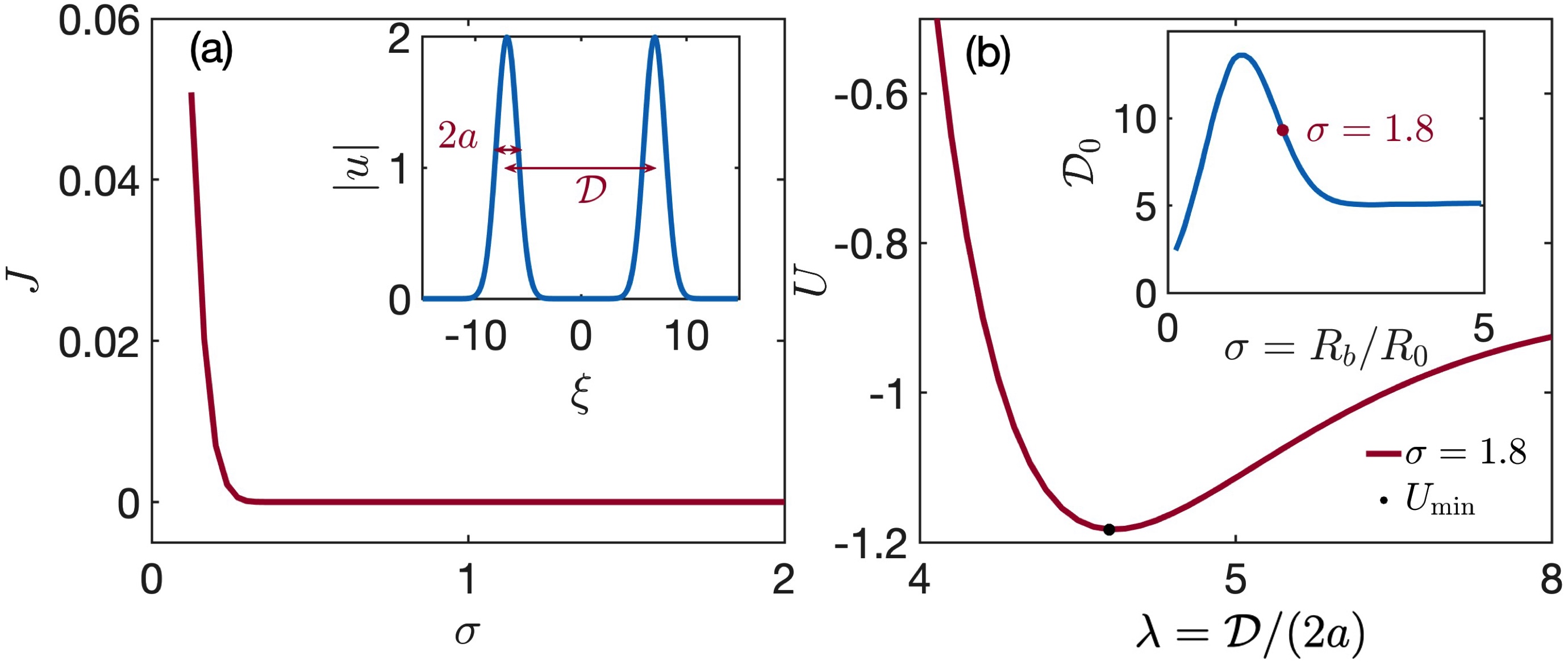}
\caption{{\protect\footnotesize Contactless interaction between two-soliton molecule.
(a)~The overlap measure $J$, defined as per Eq. (\protect\ref{J}), as a function of the nonlocality degree $\sigma$ of the Kerr nonlinearity. Inset: Amplitude $|u|$ of the two-soliton set as a function of $\protect\xi $. Inset: $|u|$ for $\eta\equiv y/R_0=0$. (b)~The effective potential $U$ of the soliton-soliton interaction vs. $\protect\lambda =\mathcal{D}/(2a) $ with  $\protect\sigma=1.8$. The black dot marks the minimum of the potential energy, $U_{\mathrm{min}}$. Inset: The equilibrium separation between the solitons, $\mathcal{D}_{0}$, vs. the nonlocality degree $\protect\sigma $.}}
\label{fig2}
\end{figure}

We recall that in locally nonlinear media, the interaction between two
solitons is determined by the overlap between their wave functions, which
quickly decays as the separation between them increases. Normally, the
interaction becomes negligible if the separation between the solitons is
$2\sim 3$ times greater than their widths~\cite{Rotschild2006}. Nevertheless, for nonlocally nonlinear media
the interaction between two solitons takes place when they have no tangible overlap, which may be called {\em contactless} interaction.
To demonstrate this clearly, we introduce an overlap parameter
\begin{equation}
J=\frac{\iint_{-\infty }^{\infty }d\xi d\eta |u_{+}u_{-}|^{2}}{%
\iint_{-\infty }^{\infty }d\xi d\eta |u_{+}|^{2}\iint_{-\infty }^{\infty
}d\xi d\eta |u_{-}|^{2}},  \label{J}
\end{equation}%
where $u_{+}$ and $u_{-}$ are introduced in Eq.~(\ref{ansatz0}).
Figure~\ref{fig2}(a)
shows $J$ as a function of $\sigma$ for the ansatz (\ref{+-}), with system's parameters $A=2$, $b=0$, $a=1$, and $\phi=0$. We see that $J$ is non-vanishing only for
$\sigma \leqslant 0.2$.

For locally nonlinear media (for which $\sigma\sim 0$), the interaction between two solitons is negligible when there is no overlap between the solitons. However, the interaction between the two solitons is non-zero even they has no overlap if the nonlocality degree of the Kerr nonlinearity reaches to a critical value [i.e.,
$\sigma \geqslant 0.2$ in the present Rydberg gas]. To characterize the size of the SM, we define the separation-width ratio
\begin{equation}
\lambda =\mathcal{D}/(2a),
\end{equation}
i.e., the ratio of the separation of the two solitons, $\mathcal{D}$, to the width of a single soliton, $2a$ [see Fig.~\ref{fig2}(a)].  Shown in Fig.~\ref{fig2}(b) is the effective interaction potential $U$\, (calculated following Ref. \cite{potential}) as a function of $\lambda $ for the ansatz (\ref{+-}) with $\sigma =1.8$, $A=2$, $b=0$, $a=1$, and $\phi=0$. It is seen that $U$ has a minimum $U_{\mathrm{min}}$ at $\lambda \approx 4.6$, which corresponds to separation $\mathcal{D}=\mathcal{D}_{0}\approx 9.2$ between the two
solitons. Thus, it is possible to expect the existence of the SM with size close to $\mathcal{D}_{0}$. Because the width of each soliton in the SM is $2a=2$, and the separation between them at the equilibrium position is $\mathcal{D}_{0}=9.2$, the effective soliton interaction is indeed \textit{contactless}. The reason for the occurrence of such a contactless interaction between the solitons
is due to the giant nonlocal Kerr nonlinearity, which induces significant interaction between the solitons while their wave functions exhibit no overlap in space.
On the other hand, in media with the local Kerr optical nonlinearity, the interaction between solitons, and hence the formation of SMs, is determined by the overlap of ``tails'' of the wave functions of adjacent solitons, and becomes negligible when there is no overlap between the solitons. Consequently, SMs in locally nonlinear media usually have only a small separation-width ratio.

Shown in the inset to Fig.~\ref{fig2}(b) is the equilibrium separation $%
\mathcal{D}_{0}$ between the two solitons as a function of the degree of the
nonlocality of the Kerr nonlinearity $\sigma $. It is seen that, as $\sigma $
increases, $\mathcal{D}_{0}$ grows at first, reaches its maximum, and
decreases, eventually saturating to a small value. Thus, by changing $\sigma
$, one can control $\mathcal{D}_{0}$ and thus the size of the SM. The
maximum value of $\mathcal{D}_{0}$ is obtained at $\sigma \approx 1.6$, which is about 13.6, and the corresponding size of
the SM in physical units is $13.6\,R_{0}\approx82$ $\mu$m. Such a value is a realistic one for the experimental observation in Rydberg gases~\cite{Bloch}.

\subsection{The formation and propagation of nonlocal (2+1)D spatial
soliton molecules}

\subsubsection{Two-soliton molecules}

\begin{figure}[ht]
\centering
\includegraphics[width=0.6\columnwidth]{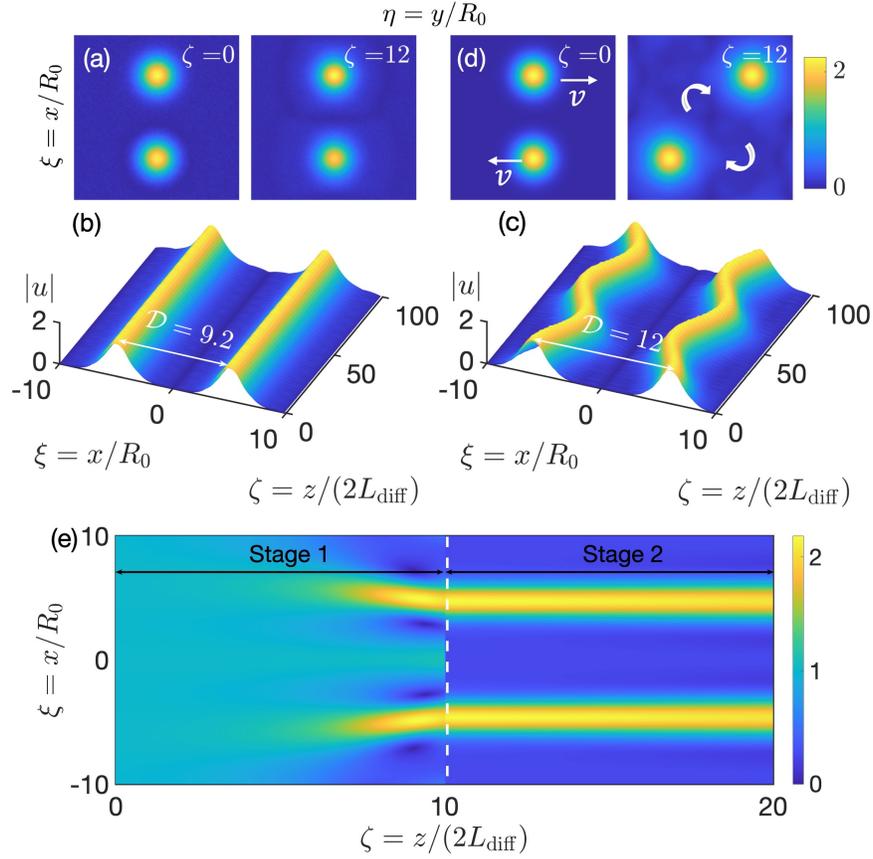}
\caption{{\protect\footnotesize The propagation of a (2+1)D spatial two-soliton molecule.
(a)~Amplitude profiles $|u|$, at different propagation distances, $\zeta =z/(2L_{\mathrm{diff}})=0$ and$\,12$. When two solitons are
placed in their equilibrium positions, they form a stable molecule. Here $\protect\sigma =1.8$, $A=2$, $a=1$, $b=0$, $\phi=0$, and $\mathcal{D}=\mathcal{D}_{0}=9.2$.
(b)~When the solitons are placed in their equilibrium positions, they form a stable (2+1)D spatial SM. Parameters are the same as in (a).
(c)~When the solitons are initially shifted from their equilibrium positions, they perform a small oscillation around the equilibria, if no additional perturbations are introduced. Parameters are the same as (a).
(d)~Amplitude profiles $|u|$ with nonzero initial velocity $v=\pm0.5$ at propagation distances $\protect\zeta =z/(2L_{\mathrm{diff}})=0\ $and$\,12$. In that case, white arrows designate the direction of the rotation of the emerging SM.
(e) The buildup of stable
(2+1)D spatial SMs. It display the first ($0\leqslant
\protect\zeta \leqslant 10$) and second stages ($10\leqslant \protect\zeta \leqslant 20$) of the evolution. Shown in
(b), (c), (e)  are for $|u|$ in the cross section $\eta\equiv y/R_0=0$.
}}
\label{fig3}
\end{figure}

We proceed to the investigation of the formation and propagation of stable
(2+1)D spatial SMs by means of numerical simulations of Eq. (\ref{NLS1}). Figure ~\ref{fig3}(a)
shows the amplitude of a typical two-soliton molecule at $\zeta =0$ and $12$, the latter value corresponding, in physical units, to
$z\approx1.2$ cm for $L_{\mathrm{diff}}\approx 0.49$ mm. The initial condition for the simulation are chosen as per ansatz (\ref{+-}) with a small random perturbation introduced by the factor
\begin{equation}
{\rm Random}=1+\epsilon R(\xi,\eta), \label{R}
\end{equation}
multiplying the initial configuration. Here, $\epsilon \ll 1$ is the amplitude of the perturbation, and $R$ is a random variable uniformly distributed in the interval [$0,1$]. Parameters of the initial condition are
taken as $\sigma =1.8$, $A=2$, $a=1$, $\mathcal{D}=\mathcal{D}_{0}=9.2$ ($\lambda \approx 4.6$), $b=\phi=0$, and $\epsilon =0.05$. The SM is found to be stable as it relaxes to the self-cleaned form close to the unperturbed one and undergoes no apparent distortion during propagation.

Shown in Fig.~\ref{fig3}(b) is the case where the two solitons are initially placed in their equilibrium positions. We see that the SM is stable and without conspicuous intrinsic oscillations. However, when the two solitons are slightly shifted from their equilibrium positions (by taking $\mathcal{D}=12>\mathcal{D}_{0}=9.2$ at $\zeta =0$), they perform a small oscillation around the equilibria, if no additional perturbations are introduced; see Fig.~\ref{fig3}(c). The existence of such a {\it excited state} of the SM  clearly
corroborates that the static copropagation of the two solitons in Fig. \ref%
{fig3}(a) [as well as in Figs. \ref{fig3}(e), \ref{fig5}(a) and \ref{fig6}
below] is indeed provided by the fact that they form the stable bound state,
rather than by trivial absence of interaction between two well-separated
solitons.

Figure~\ref{fig3}(d) illustrates the same outcome of the evolution, except that we have added initial velocities $v=\pm 0.5$ to the solitons, to admit the consideration of the generation of the two-soliton molecule under experimentally relevant conditions, and the study of the role of the incident velocity of the solitons, see~\cite{note104}). In this case, the SM exhibits persistent rotation, keeping its stability in the course of the propagation. As the rotation results in an additional centrifugal force acting on each soliton, the size of the rotating SM is larger than that of the non-rotating one.

Another way for spontaneous generation of (2+1)D spatial SMs is provided by
the modulational instability (MI). Figure~\ref{fig3}(e) shows the MI-driven
buildup of stable (2+1)D SMs. The initial condition for the simulations is
chosen as a slightly perturbed plane-wave state, i.e., $1+\epsilon \cos
(0.5\xi )$ with $\epsilon = 10^{-4}$. It is seen that, at the first stage
of the evolution, $0\leqslant \zeta \leqslant 10$, the MI develops and two bright
solitons appears, which bind into a soliton molecule along with some small radiations. At the second stage, $10\leqslant \zeta \leqslant 20$, we propagate the emerging SM after filtering out small radiation. It is found that the SM, built at the first stage of the evolution, remains stable over a very long distance.

The input power used for the generation of (2+1)D spatial SMs considered here
can be estimated by computing the corresponding Poynting's vector
integrated over the cross-sectional area of the probe beam, i.e.,
$P=\int dS (\mathbf{E}_{p}\times \mathbf{H}_{p})\cdot \mathbf{e}_z$,
where $\mathbf{e}_z$ is the unit vector in the propagation direction. Assuming that $\mathbf{E}_{p}=(E_{p},0,0)$ and $\mathbf{H}_{p}=(0,H_{p},0)$, with $H_{p}=\varepsilon_0cn_{p}E_{p}$ ($n_{p}$ is the refractive index), one can readily obtain
\begin{equation}
P_{\mathrm{gen}}=2\varepsilon_0cn_{p}S_0\left(\frac{2\hbar}{p_{13}}\right)^2|\Omega_{p}|^2\approx 3.6\,\mu {\rm W},
\end{equation}%
where $S_0$ denotes the cross-sectional area of the probe beam.
Thus, very low input power is sufficient for the creation of such nonlocal
(2+1)D spatial two-soliton molecules with the help of the
giant nonlocal Kerr nonlinearity in the present system. This fact may be
highly beneficial for applications to optical data procession and
transmission, using low-level light powers.

\subsubsection{Multi-soliton molecules}

Besides the two-soliton SMs, the Rydberg-EIT system can also support stable
nonlocal (2+1)D $N$-soliton molecules with $N\geqslant3$. The $N$-soliton
molecule in a ring-shape configuration (i.e., as a
\textit{soliton necklace}, which may be readily supported by nonlocal nonlinearities \cite{necklace}) can be sought by using the trial solution
\begin{equation}
u=A\sum_{n=1}^{N}e^{- [(\xi -\xi _{n})^{2}+(\eta -\eta
_{n})^{2}]/(2a^{2})+ib(\xi ^{2}+\eta ^{2})+i(\phi _{n}+\phi)} ,
\label{ansatz1}
\end{equation}
where $(\xi _{n},\eta _{n})=\rho _{0}[\cos(2\pi n/N),\sin (2\pi n/N)]$ is the center-of-mass position of the $n$-th Gaussian beam, with $\rho _{0}$ being the ring's radius. The phase of the $n$th beam is $\phi _{n}=2\pi m n/N$; the overall phase imposed on the ring-shaped configuration is $2\pi m$. Here $m$ is a positive integer taking in the interval ($N/4,N/2$]. In this way, the phase difference between two adjacent beams, $\phi_{n+1}-\phi_{n}$, is given by
$\pi/2<\phi_{n+1}-\phi_{n}/N\leqslant\pi$, which introduces a repulsive interaction between the beams and balances the attractive interaction due to the self-focusing nonlocal Kerr nonlinearity, and hence is in favour of the formation of a $N$-soliton molecule. The meanings of parameters $A$, $a$, $b$, and $\phi$ in ansatz (\ref{ansatz1}) is the same as in Eq.~(\ref{ansatz0}). Following the variational procedure similar to that used above, one can derive equations for parameters $A$, $\rho _{0}$, $b$, and $\phi$. As the number of the variational parameters for $N$-soliton molecules is much larger than for the two-soliton ones, we resort to numerical methods for solving the variational equations.

\begin{figure}[tbp]
\centering
\includegraphics[width=0.8\columnwidth]{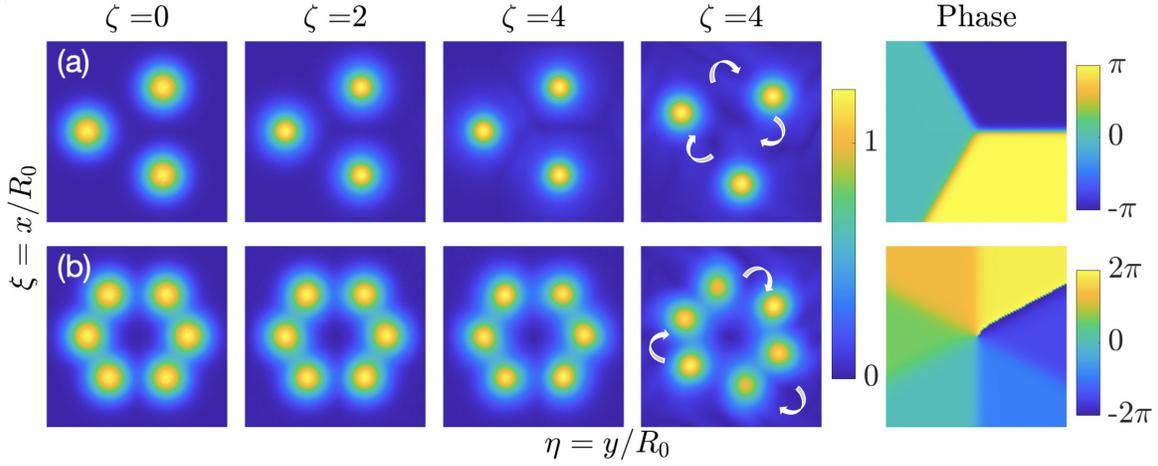}
\caption{{\protect\footnotesize The simulated propagation of a (2+1)D
necklace-shaped SM built of $N$ solitons with $N=3$ and $6$ . (a) The amplitude profile of $|u(\protect\xi ,\protect\eta )|$ with $(N, m)=(3,1)$
in the plane of $\protect\xi=x/R_{0}$ and $\protect\eta =y/R_{0}$ at different propagation distances $\protect\zeta =z/(2L_{\mathrm{diff}})=0,\,2,\,4$, respectively. The fourth column are amplitude profiles $|u|$ with nonzero initial tangential velocity $v=0.6$ at propagation distances $\protect\zeta =z/(2L_{\mathrm{diff}})=4$. In that case, white arrows designate the direction of the rotation of the emerging SM. The rightmost panel shows the phase distribution in the (2+1)D spatial SM at $\protect\zeta =0$. The parameters are $A=1.2$, $\protect\rho _{0}=5$, $a=1.45 $, and $b=0$. (b) The same as in (a), but for $(N, m)=(6,2)$.}}
\label{fig4}
\end{figure}

Fig.~\ref{fig4}(a) and Fig.~\ref{fig4}(b)
display, respectively, the simulated propagation of three- and six-soliton molecules. The initial conditions are chosen as per Eq.~(\ref{ansatz1}), respectively with $(N, m)=(3,1)$ and $(N,m)=(6,2)$, multiplied by the random-perturbation factor (\ref{R}). The parameters of the input are taken as $A=1.2$, $\rho _{0}=5$, $a=1.45$, $b=\phi =0$, and $\epsilon =0.05$. We find that both $3$- and $6$-soliton molecules are stable, relaxing to self-cleaned forms close to the unperturbed ones. The fourth column shows the same outcome of the evolution, except that we have added initial tangential velocities $v=0.6$ to the solitons, hence the SM
exhibits persistent rotation, keeping its stability in the course of the
propagation. The rightmost panel shows the initial phase distribution in the (2+1)D spatial SM.

\subsection{Nonlocal (2+1)D vortex molecules}

Stable (2+1)D spatial SMs may also be composed of vortex solitons. As an example, we consider a bound
state of two vortex solitons built as per ansatz (\ref{+-}), with each component $u_{\pm }$ taken as the
Laguerre-Gaussian beam
\begin{equation}
u_{\pm }={A}\left( \frac{\sqrt{2}r_{\pm }}{a}\right)
^{|l|}e^{-r_{\pm }^{2}/a^{2}}L_{p}^{|l|}\left( \frac{2r_{\pm }^{2}}{a^{2}}%
\right) e^{il\varphi }.  \label{ansatz2}
\end{equation}%
Here $A$ and $a$ are the soliton's amplitude and radius, $r_{\pm }=\sqrt{%
(\xi \pm \mathcal{D}/2)^{2}+\eta ^{2}}$, $L_{p}^{|l|}$ is the generalized
Laguerre-Gaussian polynomial, with the azimuthal (radial) index $l$ ($p$),
and $\varphi $ is the azimuthal angle. The ansatz based on Eqs. (\ref{+-})
and (\ref{ansatz2})~introduces the superposition of two Laguerre-Gaussian
beams with identical shapes, opposite signs, and centers placed at points $(\pm \mathcal{D}/2,0)$.
\begin{figure}[h]
\centering
\includegraphics[width=0.8\columnwidth]{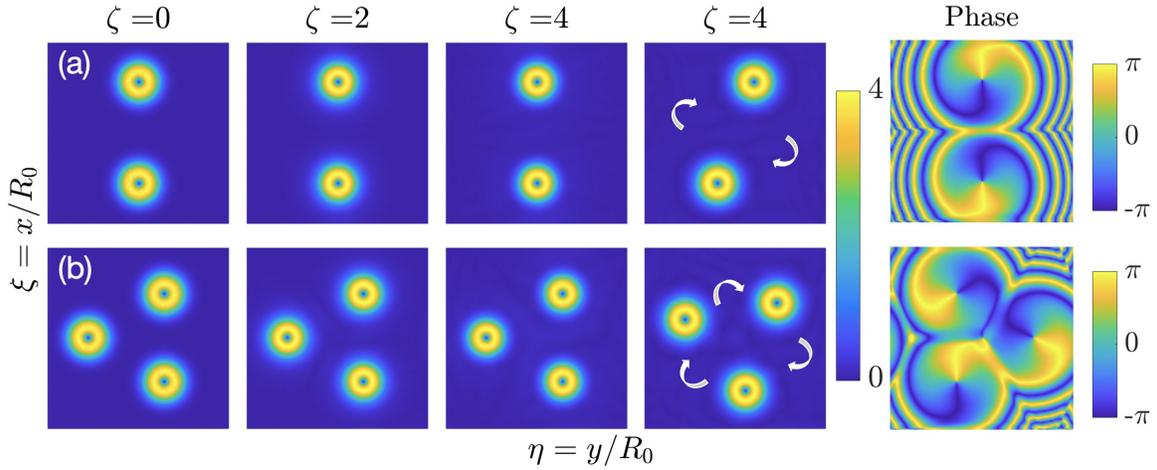}
\caption{{\protect\footnotesize The propagation of a (2+1)D vortex molecule with built of a two (a) or three (b) vortices. (a)~The amplitude profile of $|u(\protect\xi ,\protect\eta )|$ for the two-vortex SM in the plane of
$\left(\xi=x/R_{0},\,\eta =y/R_{0}\right)$ at different propagation distances $\protect\zeta=z/(2L_{\mathrm{diff}})=0,\,2,\,4$.
The fourth column are amplitude profiles $|u|$ with nonzero initial tangential
velocity $v=0.6$ at propagation distances $\protect\zeta=z/(2L_{\mathrm{diff}})=4$. In that case, white arrows designate the direction of the rotation of the emerging SM. The rightmost panel shows the phase distribution in the (2+1)D spatial SM at $\protect\zeta =2$. The azimuthal and the radial indices are $l=1$ and $p=0$ for the Laguerre-Gaussian polynomial, see
Eq.~(\protect\ref{ansatz2}). Other parameters are $A=3.9$, $a=1.45$, and $\mathcal{D}=10$.  (b) The same as in (a), but for $N=3$.
}}
\label{fig5}
\end{figure}
Figure~\ref{fig5}(a) [see also Fig.~\ref{fig1}(b)] shows the propagation of a typical (2+1)D two-vortex
molecule. Here, we fix $l=1$ and $p=0$ in the
Laguerre-Gaussian polynomial in Eq. (\ref{ansatz2}). The input, composed as
per Eqs.~(\ref{+-}) and (\ref{ansatz2}), includes the perturbation factor
(\ref{R}) too. Parameters of the input are $A=3.9$, $a=1.45$, $\mathcal{D}=10$,
and $\epsilon =0.05$. The two-vortex molecule has a large size, with the equilibrium
separation between pivots of the two vortices $\mathcal{D}_{0}=10$, corresponding to 60 $\mu\mathrm{ m}$ in physical units (in experiments with BEC, ``large" is usually a size which is essentially larger than 10 $\mu\mathrm{ m}$ \cite{Hulet2003}). Such a contactless interaction between the two vortices and the formation of the vortex molecule is also due to the nonlocal Kerr nonlinearity contributed by the long-range Rydberg-Rydberg interaction between the atoms.

We have also simulated the propagation of a three-vortex molecule, as shown
in Fig.~\ref{fig5}(b). As well as its two-vortex counterpart, it is found to be stable. The fourth column of Fig.~\ref{fig5}(a) and (b) show the same outcome of the evolution, except that we have added initial tangential velocities $v=0.6$ to the vortex, hence the vortex molecule exhibits persistent rotation, keeping its stability in the course of the propagation. The rightmost panel shows the phase distribution in the (2+1)D spatial vortex molecules at $\zeta=2$.

\section{Nonlocal (3+1)D soliton and vortex molecules, their storage and retrieval}

\subsection{Nonlocal (3+1)D soliton and vortex molecules}

The realization of (3+1)D~\cite{note102} spatiotemporal solitons is a
long-standing challenging goal of optical physics \cite{spatiotemp,Special-topics,Astra}. As mentioned above (3+1)D spatiotemporal solitons are
strongly unstable in conventional optical media with the local Kerr
nonlinearity. In a recent work, it has been shown that stable (3+1)D
spatiotemporal solitons may exist in a cold Rydberg atomic gas, being
supported by a two-step self-trapping mechanism \cite{Bai2019}.

To proceed, we first demonstrate that stable (3+1)D spatiotemporal
soliton/vortex molecules are available in the Rydberg atomic gas. To form
such states, dispersion of the probe field is necessary, which can be
secured by using a probe pulse with a short time duration. To this end, we
adopt a new set of system's parameters: $\mathcal{N}%
_{a}=10^{11}$ cm$^{-3}$, $\Delta _{2}=-2\pi \times 240$ MHz, $\Delta
_{3}=-2\pi \times 0.03$ MHz, $\Omega _{c}=2\pi \times 8$ MHz, and $\tau
_{0}=0.1\,\mu \mathrm{s}$. With these parameters, the scaled coefficients in Eq.~(\ref{NLS1}) are $d\approx 0.19$ and $w\approx 0.25$. The (3+1)D
spatiotemporal two-soliton molecules can be sought by using ansatz~(\ref%
{ansatz0}) for a single soliton, multiplying it by the temporal-localization
factor, $\mathrm{sech}(\tau /a_{\tau })\exp \left( ib_{\tau }\tau ^{2}\right) $,
where $a_{\tau }$ and $b_{\tau }$ stand for the temporal width and chirp of the probe pulse.

\begin{figure}[tbp]
\centering
\includegraphics[width=0.55\columnwidth]{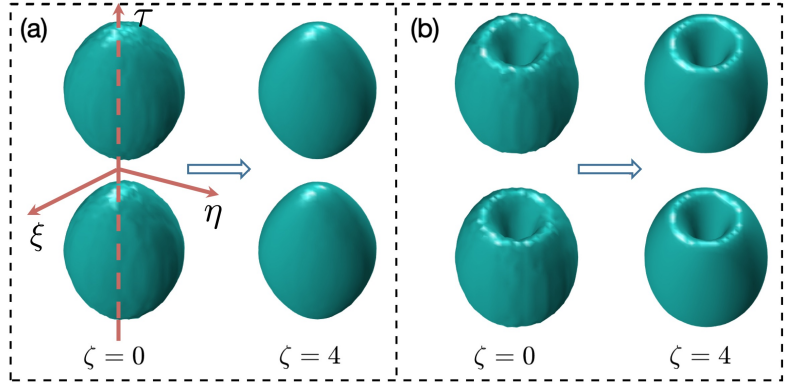}
\caption{{\protect\footnotesize The simulated propagation of
a (3+1)D spatial-temporal two-soliton molecule (a) and two-vortex molecule
(b). (a) Isosurfaces of $|u|$ of the (3+1)D
two-soliton molecule at propagation distances $\protect\zeta =z/(2L_{\mathrm{%
diff}})=~0\ $and $4$, respectively. The parameters are $A=1.6$, $a=1.2$, $\mathcal{D}=8$, $a_{\protect\tau }=1$, and $b=b_{\protect\tau }=0$.
(b)~The same as in (a), but for a (3+1)D spatial-temporal two-vortex molecule,
with the azimuthal and the radial indices $l=1$ and $p=0$ in Eq. (\protect
\ref{ansatz2}). The parameters are $A=2.5$, $a=1.3$, $\mathcal{D}=8$, $a_{\protect\tau }=1$, and $b=b_{\protect\tau }=0$.
}}
\label{fig6}
\end{figure}

Shown in Fig.~\ref{fig6}(a)
is the propagation of a typical (3+1)D spatiotemporal two-soliton molecule.
The initial condition used in the numerical simulation again includes the small random perturbation.
Parameters of the input are $A=1.6$, $\mathcal{D}=8$, $a=1.2$, $a_{\tau }=1$, $b=b_{\tau }=\phi =0$, and $\epsilon =0.05$. The (3+1)D spatiotemporal SM is found to be stable in the course of the propagation.

With the parameters given above, the propagation velocity of the (3+1)D
spatiotemporal SM, produced by the formula $V_g=(\partial K/\partial \omega)^{-1}$ at $\omega=0$, is
\begin{equation}
V_{g}=\left\{\frac{1}{c}+\kappa_{12}\frac{|\Omega_c|^2+(\omega+d_{31})^2}{[|\Omega _{c}|^{2}-(\omega +d_{21})(\omega +d_{31})]^2} \right\}^{-1}\approx3.4\times 10^{-5}\,c,
\end{equation}%
and the required generation power is estimated to be $P_{\mathrm{gen}}=6.8\,%
\mu\mathrm{W}$. Thus, the SM travels indeed with an ultraslow velocity (in
comparison to $c$) and may be created by a very low power, which is due to the
interplay of the EIT and giant nonlocal Kerr nonlinearity induced by the
Rydberg-Rydberg interaction between the atoms.

We have also carried out a numerical simulation for the propagation of a nonlocal
(3+1)D spatiotemporal two-vortex molecule, with individual vortex solitons
taken as per ansatz~(\ref{ansatz2}) times $\mathrm{sech}(\tau /a_{\tau
})\exp \left( ib_{\tau }\tau ^{2}\right) $. Fig.~\ref{fig6}(b) shows
the propagation of a typical two-vortex SM with small perturbations.
The parameters used in the simulation are $A=2.5$, $a=1.3$, $a_{\tau }=1$,
$\mathcal{D}=8$, $b=b_{\tau }=\phi =0$, and $\epsilon =0.05$. The two-vortex SM is also found to be quite stable, as well as the zero-vorticity spatiotemporal SM.

\subsection{Storage and retrieval of nonlocal (3+1)D soliton and vortex
molecules}

Keeping memory of optical pulses in atomic gases (i.e., storage of incident
pulses, with ability to retrieve them), provided by the EIT technique, has
attracted much interest \cite%
{Fleischhauer2000,Liu2001,Phillips2001,Davidson,Novikova2012,Chen2013,
Maxwell2013,Dudin2013,Heinze2013,Chen20141, Sibalic2016,Hsiao2018}. Here we
demonstrate that the storage and retrieval of (3+1)D spatiotemporal
soliton/vortex molecules are possible in the present Rydberg-EIT system. To
this end, we investigate the evolution of the (3+1)D spatiotemporal
molecules, considered above, by solving the MB Eqs.~(\ref{Bloch0}%
) and (\ref{Max}) numerically, using a time-dependent control field
\begin{equation}
\Omega _{c}(t)=\Omega _{c0}\left[ 1-\frac{1}{2}\tanh \left(\frac{t-T_{\mathrm{off}}%
}{T_{s}}\right)+\frac{1}{2}\tanh \left(\frac{t-T_{\mathrm{on}}}{T_{s}}\right)\right],
\label{Omega}
\end{equation}
which provides switching action for the probe field. Here
$T_{\mathrm{off}}$ and $T_{\mathrm{on}}$ are, respectively, the times at which the control field is switched off and on. The switching duration is $T_{s}$ and the storage time is $T_{\mathrm{on}}-T_{\mathrm{off}}$. Importantly, the switching speed of the control laser has a marginal effect on the quality of the SM storage and retrieval. This is because the (3+1)D spatiotemporal SM in the present system propagates with an ultraslow velocity, due to the EIT effect~\cite{Scully2001}.

\begin{figure*}[ht]
\centering
\includegraphics[width=1.0\textwidth]{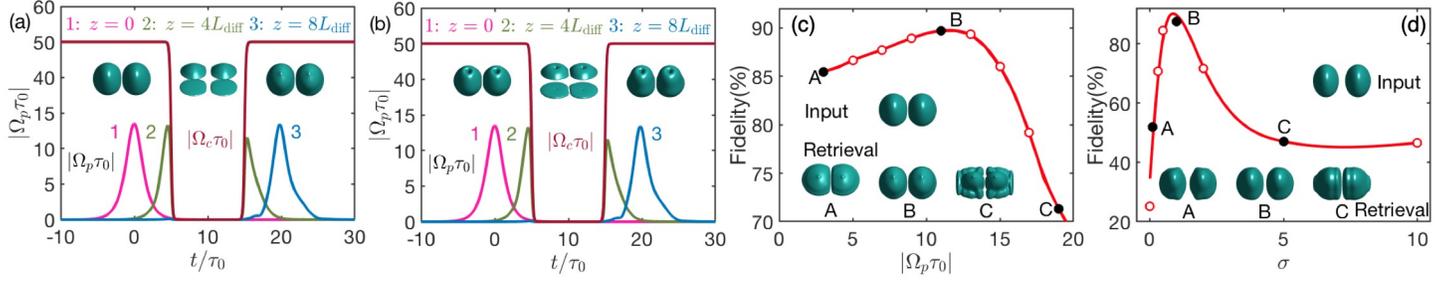}
\caption{{\protect\footnotesize Storage and retrieval of (3+1)D two-soliton
molecule and two-vortex molecule. (a)~The
storage and retrieval of a nonlocal (3+1)D spatial-temporal two-soliton
molecule. The red solid line shows switching the control field $|\Omega_{c}%
\protect\tau _{0}|$ on and off. Curves 1, 2, and 3 are temporal profiles of
the probe pulse $|\Omega_{p}\protect\tau _{0}|$, respectively, at $z=0$
(the initial condition), $z=4L_{\mathrm{diff}}$ (just before the storage),
and $8L_{\mathrm{diff}}$ (just after the retrieval), with $L_{\mathrm{diff}%
}=0.87$ mm; the corresponding isosurface plots for $|\Omega _{p}\protect\tau %
_{0}|=0.5$ are shown. (b)~The same as (a) but for the storage and retrieval
a nonlocal (3+1)D spatial-temporal two-vortex molecule. (c)~The fidelity $%
\protect\eta {\mathcal J}$ as a function of the probe-field amplitude $|\Omega _{p}%
\protect\tau _{0}|$ at $\protect\zeta =z/(2L_{\mathrm{diff}})=4$. The
isosurfaces of the input probe field (upper inset) and the retrieved ones at
different $|\Omega _{p}\protect\tau _{0}|$ (lower insets) are also
illustrated. (d) The fidelity $\protect\eta {\mathcal J}$ as a function of the
nonlocality degree $\protect\sigma $ at $\protect\zeta =z/(2L_{\mathrm{diff}%
})=4$. The isosurfaces of the input probe field (upper inset) and the
retrieved ones at different $\protect\sigma $ (lower insets) are also
illustrated. }}
\label{fig7}
\end{figure*}

Shown in Fig.~\ref{fig7}(a)
is the evolution of the probe-pulse amplitude $%
|\Omega _{p}\tau _{0}|$ in the course of the storage and retrieval process.
The shapes of the probe pulse at $z=0$ (before storage), $z=4L_{\mathrm{diff}%
}$ (at the beginning of storage), and $z=8L_{\mathrm{diff}}$ (after the
retrieval) are shown for $L_{\mathrm{diff}}=0.87$ mm. It is seen that
switching off the control field provides for the storage of the (3+1)D
spatiotemporal SM in the atomic medium, which is retrieved when the control
field is switched on again. Further, the retrieved spatiotemporal SM has
nearly the same shape as the original one prior to the storage. In the
course of the storage, the information carried by the SM is converted into
that kept in the atomic spin wave (i.e., the coherence matrix element $\rho
_{13}$). A slight deformation affecting the optical memory is due to
dissipation, including spontaneous emission and dephasing, as well as weak
imbalance between diffraction, dispersion, and nonlinearity. We have also
explored the storage and retrieval of (3+1)D spatiotemporal vortex
molecules. Similar results are obtained for the vortex bound states, as
shown in Fig.~\ref{fig7}(b).

The quality of the storage and retrieval of nonlocal (3+1)D spatiotemporal
SMs can be characterized by
efficiency $\eta $ and fidelity $\eta {\mathcal J}$, where $\eta $ and ${\mathcal J}$ are defined
as
\begin{subequations}
\begin{align}
\eta & =\frac{\int_{T_{\mathrm{on}}}^{\infty }dt\iint dxdy|\Omega _{p}^{%
\mathrm{out}}(x,y,t)|^{2}}{\int_{-\infty }^{T_{\mathrm{off}}}dt\iint
dxdy|\Omega _{p}^{\mathrm{in}}(x,y,t)|^{2}}, \\
{\mathcal J}& =\frac{|\int_{-\infty }^{\infty }dt\iint dxdy\Omega _{p}^{%
\mathrm{out}}(x,y,t-\Delta T)\Omega _{p}^{\mathrm{in}}(x,y,t)|^{2}}{%
\int_{-\infty }^{T_{\mathrm{off}}}dt\iint dxdy|\Omega _{p}^{\mathrm{in}%
}|^{2}\int_{T_{\mathrm{on}}}^{\infty }dt\iint dxdy|\Omega _{p}^{\mathrm{out}%
}|^{2}}.
\end{align}%
\end{subequations}
Based on the results obtained in Fig.~\ref{fig7}(a) and (b), we obtain $\eta
=90.39\%$, ${\mathcal J}=98.81\%$, and $\eta {\mathcal J}=89.31\%$ for the (3+1)D spatiotemporal
SM, and $\eta =90.32\%$, ${\mathcal J}=97.48\%$, and $\eta {\mathcal J}=88.05\%$ for the vortex
molecule.

The strength of the nonlocal Kerr nonlinearity has a significant effect on
the quality of the optical memory. Figure~\ref{fig7}(c) shows fidelity $\eta
{\mathcal J}$ of the retrieved (3+1)D spatiotemporal SM as a function of the
probe-pulse amplitude. For this purpose, a set of probe-pulse isosurfaces, $%
|\Omega _{p}\tau _{0}|=3$, $11$, and $19$ at $z=8L_{\mathrm{diff}%
}\approx 7$ mm are displayed. For the moderate amplitude, $|\Omega _{p}\tau
_{0}|\approx 11$, the fidelity reaches its maximum, with the retrieved
spatiotemporal SM having nearly the same shape as the original one prior to
the storage. For small and large amplitudes, the fidelity features small
values, i.e., the retrieved SM is distorted greatly. This happens because,
for the weak and strong probe-pulse amplitudes, the Kerr nonlinearity is
either too weak or too strong to balance the diffraction and dispersion.

The nonlocality degree of the Kerr nonlinearity also has a significant
effect on the memory quality for the spatiotemporal SMs. Figure~\ref{fig7}%
(d) shows fidelity $\eta {\mathcal J}$ of the retrieved (3+1)D SM as a function of the
nonlocality degree, $\sigma $. The probe-pulse isosurfaces ($|\Omega
_{p}\tau _{0}|=0.5$) are displayed for $\sigma =0.5$, $1.4$, and $5.0$ at $%
z=8L_{\mathrm{diff}}\approx 7$ mm. The fidelity reaches its maximum in the
case of the moderate nonlocality degree, $\sigma \approx 1.4$, letting the
retrieved SM keep nearly the same shape as the original one had. In the
cases of small and large nonlocality degrees, the fidelity may have only
small values, greatly distorting the retrieved SM. This happens because, in
the limit of local response ($\sigma \rightarrow 0$), the Kerr nonlinearity
becomes local, making all (3+1)D solitons unstable, as mentioned above. On
the other hand, in the limit of the strongly nonlocal response ($\sigma
\rightarrow \infty $), the nonlocal Kerr nonlinearity reduces to a linear
potential (as in the above-mentioned \textquotedblleft accessible-soliton"
model \cite{Snyder1997}), which cannot support stable (3+1)D SMs either.

\section{Conclusion}

We have elaborated a scheme which makes it possible to create stable optical
multidimensional SMs (soliton molecules), i.e., bound states of
zero-vorticity solitons, as well as bound states of vortex solitons, in a
gas of cold Rydberg atoms, in which the laser illumination maintains the EIT
setting. Due to the interplay of EIT and strong long-range Rydberg-Rydberg
interaction between atoms, the system gives rise to giant nonlocal Kerr
nonlinearity, which provides for the stability of the (2+1)-dimensional SMs
(that would be completely unstable under the action of the local
nonlinearity). The SMs feature large sizes, low generation powers, and can
be effectively controlled by means of the nonlocality degree of the Kerr
nonlinearity. The system allows, as well, the creation of stable (3+1)D
spatiotemporal SMs, including those built of vortex spatiotemporal solitons,
moving with ultraslow velocities and requiring very low generation powers.
Further, the spatiotemporal solitons can be stored and retrieved through the switching off and on of the control laser field. The findings reported here provide insight into the use of long-range atomic interactions for creating robust
bound states of solitons and developing methods to effectively control them.
The predictions reported here are helpful for experimental observations of  high-dimensional soliton molecules and promising to find applications to optical data processing and transmission.

\medskip
\textbf{Supporting Information} \par
Supporting Information is available from the Wiley Online Library or from the author.

\medskip
\textbf{Acknowledgements} \par 
We thank J. Peng for useful discussions. G. H. and C. H. acknowledge the support from National Natural Science Foundation of China (11975098, 11974117); C. H. acknowledges the support from National Key Research and Development Program of China (Nos. 2016YFA0302103 and 2017YFA0304201); Shanghai Municipal Science and Technology Major Project (No. 2019SHZDZX01); B. A. M. acknowledges the support Israel Science Foundation (1286/17).
\medskip

\medskip
\textbf{Conflict of Interest} \par
The authors declare no conflict of interest.

\bibliographystyle{MSP}
\bibliography{references}

\begin{thebibliography}{100}
\bibitem{note100}
 Here $k_{1}=0$, $k_{2}=k_{p}$, and $k_{3}=k_{p}+k_{c}$. In the Schr\"{o}dinger
  picture, $\hat{\sigma}_{\alpha \beta }(z)\equiv |\alpha \rangle
  _{z\,z}\langle \beta |$, where $|\alpha \rangle _{z}$ is energy eigenstate
  $|\alpha \rangle $ of the atom located at position $z$.
\bibitem{note104}
 Such soliton molecules can be generated experimentally. For example, to obtain
  a 2D spatial soliton molecule consisting of two spatial solitons, one needs
  firstly to prepare two solitons at the entrance of the Rydberg gas. Then, the
  solitons are incident to the Rydberg gas with a proper space separation and
  initial velocities. The initial velocity of each soliton is acquired when the
  soliton is incident obliquely. The relation between the soliton velocity $v$
  and the incidence angle $\theta$ is given by $\theta={\rm arctan}\,(v)$.
\end{thebibliography}

\end{document}